# The Red Radio Ring: a gravitationally lensed hyperluminous infrared radio galaxy at z=2.553 discovered through the citizen science project SPACE WARPS

J. E. Geach[1], A. More[2], A. Verma[3], P. J. Marshall[4], N. Jackson[5], P.-E. Belles[5], R. Beswick[5], E. Baeten[6], M. Chavez[7], C. Cornen[6], B. E. Cox[8], T. Erben[9], N. J. Erickson[10], S. Garrington[5], P. A. Harrison[5], K. Harrington[10], D. H. Hughes[7], R. J. Ivison[11,12], C. Jordan[5], Y.-T. Lin[13], A. Leauthaud[2], C. Lintott[3], S. Lynn[15], A. Kapadia[14], J.-P. Kneib[15,16], C. Macmillan[6], M. Makler[17], G. Miller[6], A. Montaña[7], R. Mujica[7], T. Muxlow[5], G. Narayanan[10], D. Ó Briain[18], T. O'Brien[5], M. Oguri[19], E. Paget[14], M. Parrish[14], N. P. Ross[11], E. Rozo[20,21], C. E. Rusu[22], E. S. Rykoff[4], D. Sanchez-Argüelles[7], R. Simpson[3], C. Snyder[14], F. P. Schloerb[10], M. Tecza[3], W-H. Wang[13], L. Van Waerbeke[23], J. Wilcox[6], M. Viero[4], G. W. Wilson[10], M. S. Yun[10], M. Zeballos[7]

[1]*Centre for Astrophysics Research, Science & Technology Research Institute, University of Hertfordshire, Hatfield, AL10 9AB, UK. j.geach@herts.ac.uk*
[2]*Kavli Institute for the Physics and Mathematics of the Universe, University of Tokyo, Kashiwa, Chiba 277-8582, Japan*
[3]*Oxford Astrophysics, Denys Wilkinson Building, Keble Road, Oxford, OX1 3RH, UK*
[4]*Kavli Institute for Particle Astrophysics and Cosmology, P.O. Box 20450, MS 29, Stanford, CA 94309, USA*
[5]*e-MERLIN/Jodrell Bank Centre for Astrophysics, School of Physics and Astronomy, The University of Manchester, M13 9PL, UK*
[6]*Zooniverse, c/o Oxford Astrophysics, Denys Wilkinson Building, Keble Road, Oxford, OX1 3RH, UK*
[7]*Instituto Nacional de Astrofĺ́Aĺsica, ĺAÓptica y Electrĺ́Aónica, Luis Enrique Erro #1, Tonantzintla, Puebla 72840, MĺAÉxico*
[8]*School of Physics & Astronomy, University of Manchester, Oxford Road, Manchester, M13 9PL, UK*
[9]*Argelander Institute for Astronomy, University of Bonn, Auf dem Hügel 71, D-53121 Bonn, Germany*
[10]*Department of Astronomy, University of Massachusetts, Amherst, MA 01002, USA*
[11]*Institute for Astronomy, University of Edinburgh, Royal Observatory, Blackford Hill, Edinburgh, EH9 3HJ, UK*
[12]*European Southern Observatory, Karl-Schwarzschild-Str. 2, D-85748, Garching bei München, Germany*
[13]*Institute of Astronomy and Astrophysics, Academia Sinica, Taiwan*
[14]*Adler Planetarium, 1300 South Lake Shore Drive, Chicago, IL 60605, USA*
[15]*Laboratoire d'Astrophysique, Ecole Polytechnique Fédérale de Lausanne, Observatoire de Sauverny, CH-1290 Versoix, Switzerland*
[16]*Aix Marseille UniversitÃl', CNRS, Laboratoire d'Astrophysique de Marseille, UMR 7326, 13388, Marseille, France*
[17]*Centro Brasileiro de Pesquisas FĺAísicas, Rua Dr. Xavier Sigaud 150, Rio de Janeiro, RJ 22290-180, Brazil*
[18]*c/o British Broadcasting Corporation*
[19]*Research Center for the Early Universe, University of Tokyo, 7-3-1 Hongo, Bunkyo-ku, Tokyo 113-0033, Japan*
[20]*SLAC National Accelerator Laboratory, Menlo Park, CA 94025, USA*
[21]*University of Arizona, Department of Physics, 1118 E. Fourth St., Tucson, AZ 85721, USA*
[22]*Optical and Infrared Astronomy Division, National Astronomical Observatory of Japan, 2-21-1 Osawa, Mitaka, Tokyo 181-8588, Japan*
[23]*Department of Physics and Astronomy, University of British Columbia, 6224, Agricultural Road, Vancouver, B.C., V6T 1Z1, Canada*

7 May 2015

## ABSTRACT

We report the discovery of a gravitationally lensed hyperluminous infrared galaxy (**intrinsic** $L_{\rm IR} \approx 10^{13} L_{\odot}$) with strong radio emission (**intrinsic** $L_{\rm 1.4GHz} \approx 10^{25}\,{\rm W\,Hz^{-1}}$) at $z = 2.553$. The source was identified in the citizen science project SPACE WARPS through the visual inspection of tens of thousands of $iJK_{\rm s}$ colour composite images of Luminous Red Galaxies (LRGs), groups and clusters of galaxies and quasars. Appearing as a partial Einstein ring ($r_e \approx 3''$) around an LRG at $z = 0.2$, the galaxy is extremely bright in the sub-millimetre for a cosmological source, with the thermal dust emission approaching 1 Jy at peak. The redshift of the lensed galaxy is determined through the detection of the CO(3→2) molecular emission line with the Large Millimetre Telescope's Redshift Search Receiver and through [O III] and Hα line detections in the near-infrared from Subaru/IRCS. We have resolved the radio emission with high resolution (300–400 mas) eMERLIN L-band and VLA C-band imaging. These observations are used in combination with the near-infrared imaging to construct a lens model, which indicates a lensing magnification of $\mu \approx 10$. The source reconstruction appears to support a radio morphology comprised of a compact ($< 250\,{\rm pc}$) core and more extended component, perhaps indicative of an active nucleus and jet or lobe.

# 1  INTRODUCTION

Strong gravitational lenses are striking astrophysical laboratories that allow us to study the structure and distribution of matter in and around massive galaxies and clusters (Koopmans et al. 2006, Ruff et al. 2011, Sonnenfeld et al. 2013, Richard et al. 2014), cosmography through quasar time delays (Refsdal 1964, Tewes et al. 2013, Suyu et al. 2013), perform highly detailed investigations of early galaxies (e.g. Kneib et al. 2005, Stark et al. 2008, Swinbank et al. 2009) and even detect the first galaxies (Kneib et al. 2004, Zheng et al. 2012, Zitrin et al. 2014). The number of confirmed gravitational lenses is still relatively low: only some 500 are confirmed to date. With diverse astrophysical and cosmological applications, surveys for new strong gravitationally lensed systems remain a useful endeavour. In this article, we present new results from a novel search for new gravitational lens systems using wide-field near-infrared and optical data. Lens searches in the visible bands are typically optimised for finding blue arcs, representing the images of star-forming galaxies in the early Universe (which are abundant). Searches at longer (near-infrared and beyond) wavelengths have the advantage of selecting dust-obscured or very high-redshift sources that could easily be missed in optical searches (**e.g. Negrello et al. 2010; Vieira et al. 2013**).

The VISTA-CFHT Stripe 82 survey (VICS82, Geach et al. in preparation) is a wide area near-infrared survey of the Sloan Digital Sky Survey (SDSS) Stripe 82 with CFHT/WIRCam and VISTA/VIRCAM, covering 130 square degrees to a $5\sigma$ depth of approximately 22nd AB magnitude in the $J$ and $K_s$ bands. Stripe 82 is also covered by deep (24th magnitude) $i$-band imaging from the Canada-France-Hawaii Telescope (CFHT) in the CS82 survey (Moraes et al. 2014, Erben et al. in prep) with comparable image quality (sub-arcsecond), as well as the deeper coverage from SDSS itself across the equatorial stripe. The combination of moderately deep optical-near-infrared imaging over a wide area is ideal for searches for strong, 'red' gravitational arcs. After serendipitously finding a handful of promising lens candidates through visual inspection, we decided to pass the dataset to SPACE WARPS (Marshall et al. 2015, More et al. 2015). SPACE WARPS is a citizen science project to discover gravitational lenses run as part of the *Zooniverse* family[1] of online projects (Marshall et al. 2015, More et al. 2015).

Gravitational lens discovery remains a labour-intensive process, even automated algorithms produce highly impure samples that require visual inspection for verification (Marshall et al. 2009, More et al. 2012, Gavazzi et al. 2014, Chan et al. 2014, Brault & Gavazzi 2014). SPACE WARPS was conceived to **mitigate** this issue for gravitational lens searches in the era of large surveys which have the potential to increase current sample sizes by several orders of magnitude. Through an online interface, users (the general public) are presented with tens of thousands of images, each of which are inspected for evidence of strong gravitational lensing. A sophisticated training scheme using images containing realistic simulated lenses, built into the inspection interface, is used to simultaneously train users and calibrate the classification process (Marshall et al. 2015, More et al. 2015). Humans are highly proficient at detecting the – often very subtle – features associated with gravitational lenses, and so harnessing the power of a natural neural network of several tens of thousands of 'nodes' is a particularly efficient approach to this problem.

The VICS82+CS82 search with SPACE WARPS (actually the second incarnation of the project, the first was an optical search within the CFHT-Legacy Survey) was accelerated by being launched on *Stargazing Live!*, broadcast on 7th January 2014 to approximately 2 million viewers in the United Kingdom by the British Broadcasting Corporation. The goal of *Stargazing Live!* is to engage the public with astronomy-related activities ranging from amateur pursuits through to the communication of the latest scientific research, major missions and projects. The wide reach of this programme (which was run over three consecutive nights) makes it an exceptionally successful and efficient public engagement tool for promoting astronomy, and indeed science in general. SPACE WARPS was phenomenally successful: by the end of the final programme on 9th January, over 7.5 million classifications of images had been logged, such that each $iJK_s$-band image in the sample was independently viewed approximately two hundred times. Out of several tens of excellent lens candidates identified by the collective efforts of all participants of SPACE WARPS , the best candidate we have appears to be an active, dust-obscured source at $z \approx 2.5$. Dubbed '9io9' (a shortening of its SPACE WARPS identifier ASW0009io9[2]), the target appears as an incomplete red Einstein ring of radius approximately $3''$ around a Luminous Red Galaxy (LRG) at $z = 0.2$ (Figure 1).

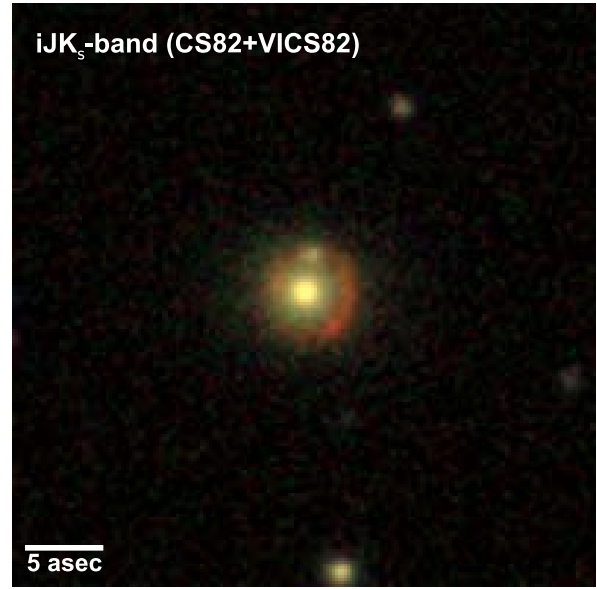

**Figure 1.** The colour composite image of 9io9 that was presented to online visitors to SPACE WARPS . Users were presented with a (random) sequence of $40''$ cut-out images where the blue channel represents $i$-band data from the CHFT Stripe 82 (CS82) survey (Erben et al. in prep) and the green and red channels are $J$ and $K_s$-band imaging from the VISTA-CFHT Stripe 82 (VICS82) survey (Geach et al. in prep). Images were centred on Luminous Red Galaxies (LRGs), groups and clusters and quasars. Volunteers were asked to identify potential lensed features, having been trained using simulated images. In this case, a red ring of radius $\approx 3''$ can be seen around the central galaxy (an LRG at $z = 0.2$). This system was independently identified by several users and quickly became a high-confidence lens candidate.

In this paper we present follow-up observations and lens modelling of 9io9 that reveal the galaxy to be **an intrinsically** radio- and sub-millimetre-bright hyperluminous infrared galaxy (HLIRG, $L_{\rm IR} > 10^{13} L_\odot$) at $z = 2.553$. In Section 2 we describe archival and new observations of 9io9, in Section 3 we constrain the integrated spectral energy distribution of the galaxy and model the lens itself, and in Section 4 we conclude with a brief discussion of the likely nature of 9io9. Throughout we assume a fiducial cosmology where $\Omega_m = 0.3$, $\Omega_\Lambda = 0.7$ and $H_0 = 70 \, \text{km s}^{-1} \, \text{Mpc}^{-1}$.

---

[1] http://zooniverse.org

[2] http://talk.spacewarps.org/#/subjects/ASW0009io9

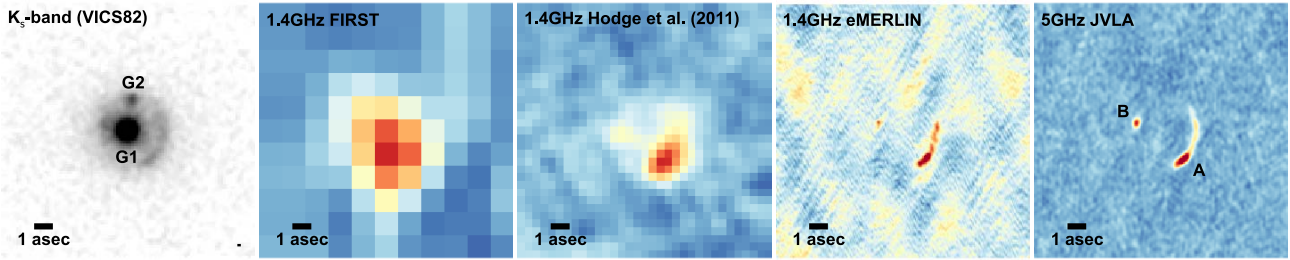

**Figure 2.** Images of 9io9 (aligned north up, east left): left-to-right: (a) $K_s$-band image. The small point source (G2) to the north of the central (lensing) galaxy G1 is not a background source, but likely a satellite of the lensing galaxy at $z = 0.202$ (b) VLA FIRST 1.4 GHz image showing the radio detection, peaking slightly SW of the central LRG, (c) higher resolution 1.4 GHz VLA imaging from the survey of Hodge et al. (2011), showing that the radio emission resolves into a ring configuration that traces the near-infrared light, but with a clear peak at the SW, (d) eMERLIN 1.4 GHz map (300 mas beam) resolving two main radio components: an arc (A) and point source (B), (e) VLA 5 GHz image at similar resolution, again resolving the radio morphology. Our lens model (§3.2) indicates that B is the counterimage of the bright, compact component at the southern extent of A.

## 2 OBSERVATIONS

### 2.1 Radio

Immediately upon the discovery of the lens candidate during SPACE WARPS we cross-checked the position of the source ($02^h09^m41.3^s$, $+00°15'58.6''$) in other imaging surveys. The Very Large Array (VLA) FIRST 1.4 GHz survey (Becker et al. 1994) reveals a 4 mJy radio source at this position, but the angular resolution of FIRST is too coarse to determine whether the radio emission originates from the background galaxy, or the lensing galaxy. Hodge et al. (2011) present higher resolution VLA (A and B configuration, $\theta = 1.8''$) 1.4 GHz imaging of Stripe 82 that partially resolves the radio flux into a ring that follows the $K_s$-band light (Figure 2). This motivated us to obtain higher resolution radio imaging of 9io9, which we describe here.

#### 2.1.1 eMERLIN L-band

We obtained eMERLIN observations over 10.25 hrs between 1415 UT 9th January 2014 to 0030 UT 10th January 2014 (during the television broadcast). The target source was interspersed with the phase reference calibrator (J0208-0047) with a cycle time of 10 minutes with 7 minutes on the target and 3 minutes on the phase reference. 30 seconds of data were flagged at the start of each scan to remove corrupted data logged during telescope drives with a resulting total time on the target source of 6.72 hours. The data were divided into 8 intermediate frequencies (IFs) with a total observing bandwidth of 512 MHz at a centre frequency of 1.5185 GHz. Data calibration was performed from calibration scans on the resolved flux density calibrator 3C286 (1331+305) and the point source calibrator OQ208 (1407+284). The flux density scale assumed for 3C286 incorporated the latest coefficients from Perley & Butler (2013) and a standard model of spatial resolution of the source as observed by eMERLIN. These were used to derive flux densities for the point source OQ208 which has a rising spectrum across the observed band (IF1: 0.815 Jy, IF8: 1.208 Jy), with a value at the central frequency of 1.0078 Jy. Estimated errors in the flux density scale are at the level of 5–10%. Residual delay corrections for the data throughout the run were derived from all three calibration sources. After performing phase and minor gain corrections derived from the phase calibrator, the target was imaged and several rounds of phase-only self-calibration applied. The data calibration and image processing was performed in AIPS. The final beam is nearly circular with FWHM 300 mas, and the map r.m.s. is 16 μJy beam$^{-1}$.

#### 2.1.2 Karl G. Jansky Very Large Array C-band

Data were obtained during 1st March 2014 using the 3-bit samplers and an advanced mode of the immensely powerful WIDAR correlator at the heart of the **Karl G. Jansky Very Large Array** (VLA, proposal 13B–458), with eight sub-bands in each of the A1/C1, A2/C2, B1/D1 and B2/D2 basebands. We used the VLA's C-band receivers with the correlator instructed to record a band comprising $1024 \times 62.5$ kHz full-polarisation channels centred on the 22.23508 MHz water line, tracking a barycentric redshift of 2.5529. We also recorded 31 wider bands to sample continuum emission, each comprising $128 \times 1$ MHz dual-polarisation channels, covering 4.05–7.96 GHz contiguously, with a few small regions of overlap. To mitigate time smearing, since we used the widest (A) configuration of the VLA, the integration time was set to 2 seconds, which resulted in a total data rate of 62 Gb hr$^{-1}$, with $26 \times 287$ seconds (roughly 2 hours) spent integrating on 9io9, $27 \times 31$ seconds on the gain calibrator, J0215−0222, and 226 seconds on 0137+331 (3C 48) to set the flux density scale. The data were calibrated using version 4.1 of CASA, starting with a pass through the initial stages of the VLA data pipeline, then flagging extensively by hand, then re-starting the pipeline where we had previously paused. CASA was also used for imaging, which required us to map and clean a number of bright, nearby sources to reduce contamination of the region around 9io9 by strong sidelobes. The resulting image of 9io9, centred at 5 GHz, has a $500 \times 400$ mas beam (PA = 7°), with an r.m.s. noise level of 3–4 μJy beam$^{-1}$.

### 2.2 Sub-millimetre and millimetre

A large area of Stripe 82 has been mapped at 250μm, 350μm and 500μm with *Herschel*, as part of the The *Herschel* Stripe 82 Survey (HerS, Viero et al. 2014). At the position of 9io9 there is bright sub-millimetre detection, with $S_{250} = 826 \pm 7$ mJy, $S_{350} =$ $912 \pm 7$ mJy, $S_{500} = 717 \pm 8$ mJy. The low resolution of *Herschel* cannot spatially resolve the emission, but the sub-millimetre colours are consistent with the emission originating from the background source at $z \sim 2$–3 (the lensing galaxy is not expected to be a significant source of sub-millimetre emission). In order to better constrain the shape of the far-infrared spectral energy distribution and determine the redshift of the lensed source, we obtained further sub-millimetre imaging and spectroscopy that we describe in the following.

### 2.2.1 *James Clerk Maxwell Telescope SCUBA-2*

We obtained a 30 minute observation of 9io9 with the SCUBA-2 camera on the James Clerk Maxwell Telescope (JCMT) on 15th January 2014 as part of Director's Discretionary Time (project M13BD12). SCUBA-2 records both 450$\mu$m and 850$\mu$m data simultaneously. We employed the DAISY scanning mode, appropriate for point sources, and the zenith opacity at 225 GHz was $\tau_{225\mathrm{GHz}} = 0.07$–0.1 over the course of the observation. The data were reduced with the dynamic iterative map maker of the SMURF package (Jenness et al. 2011, Chapin et al. 2013) and optimally filtered for detection of point sources. We applied flux conversion factors (FCFs) of 491 Jy pW$^{-1}$ and 537 Jy pW$^{-1}$ at 450$\mu$m and 850$\mu$m respectively, which are the canonical FCFs derived from observations of hundreds of standard sub-millimetre calibrators observed by SCUBA-2 since operations began (Dempsey et al. 2013). An additional correction of 10% is made to the FCFs to account for flux lost during the filtering steps. The absolute flux calibration at 850$\mu$m is accurate to within 15%.

9io9 is detected with $S_{850} = 167 \pm 4$ mJy (the JCMT beam at 850$\mu$m is 15$''$, and so this can be considered an integrated measurement). At the same position in the 450$\mu$m map, we measure $S_{450} = 217 \pm 72$ mJy, this is a factor of three lower than the *Herschel* measured flux at approximately the same wavelength. Given the ~5× finer resolution of JCMT/SCUBA-2 compared to *Herschel*/SPIRE at 500$\mu$m, it is possible that SCUBA-2 is resolving the emission into several beams, rendering some of the sub-millimetre emission undetectable. This offers a first insight into the distribution of cold dust in 9io9: higher resolution sub-millimetre mapping with a powerful interferometer such as the Atacama Large Millimeter Array (ALMA) will allow us to compare the dust and radio morphologies on comparable angular scales.

### 2.2.2 *Large Millimeter Telescope AzTEC 1.1 mm & Redshift Search Receiver*

The Large Millimeter Telescope Alfonso Serrano, hereafter LMT, is a 50m diameter millimeter-wavelength radio telescope on Volcán Sierra Negra, Mexico, at an altitude of 4600 m. LMT 'Early Science' observations were conducted towards 9io9 with AzTEC (Wilson et al. 2008) and the Redshift Search Receiver (Erickson et al. 2007), operating in the 1mm and 3mm atmospheric windows respectively. During the 'Early-Science' programme only the inner 32m diameter of the LMT primary surface was aligned and coupled to the scientific instruments.

Continuum imaging of 9io9 was first conducted at 1.1mm with the AzTEC camera on 16th January 2014 UT, and then repeated on 24th January 2014 UT due to a poor focus in the original observations. The highest quality observations however were obtained during the early evening of November 2nd 2014 UT under improved weather conditions with a 225GHz zenith opacity of $\tau_{225\mathrm{GHz}} \approx$ 0.08. The measured beam-size at 1.1mm was $8.5'' \times 7.9''$ FWHM in azimuth and elevation respectively. The AzTEC observations were made in the standard Photometry Map Mode using a Lissajous scanning pattern, and reduced using the standard AzTEC data-analysis pipeline (Scott et al. 2008). Flux calibration was performed against Neptune, assuming a 1.1mm flux density of 18.38 Jy. A short four minute AzTEC integration detected the source 9io9 with a S/N=39 and a flux density of $95.5 \pm 2.4$ mJy, with no suggestion of extended emission. The accuracy of the absolute flux calibration is estimated to be ~10%.

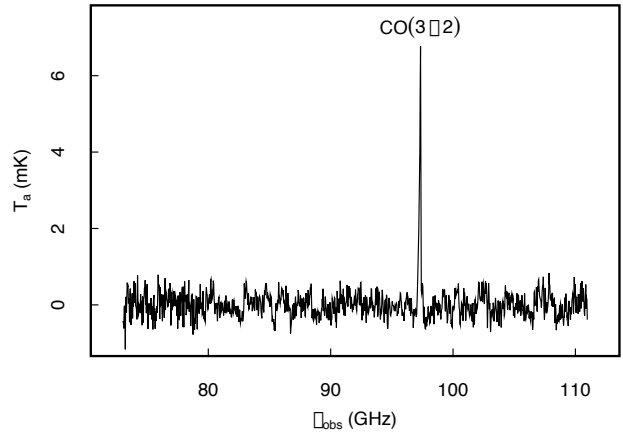

**Figure 3.** The Redshift Search Receiver spectrum of 9io9 showing the full 38 GHz instantaneous bandwidth spectral coverage in the 3 mm atmospheric window (73–111 GHz). The detected molecular line at 97.312 GHz is $^{12}$CO(3–2) from the lensed background source at a redshift $z = 2.553$.

The LMT operates with an ultra-wideband spectrometer, the Redshift Search Receiver (RSR) that covers the full 3mm atmospheric window (73–111 GHz) at low spectral resolution (31,MHz channels, ~100 km s$^{-1}$ at 90 GHz). The RSR is a dual-beam dual-polarization spectrometer, with 38 GHz instantaneous bandwidth, designed to detect molecular gas, typically CO line transitions, from high-redshift galaxies. A 1 hour integration with the RSR was conducted towards 9io9 on 19th January 2014 UT with a measured $T_{\mathrm{sys}}$ at 90 GHz of 85 K. The RSR data were reduced and calibrated using the LMT DREAMPY (Data REduction and Analysis Methods in PYthon) package. A single strong spectral-line was detected at 97.31 GHz (Figure 3). The combination of the photometric redshift of $z \sim 2.6$, determined from the (sub-)millimeter and FIR observations presented in this paper, with the RSR spectrum eliminates alternative redshift solutions and any ambiguity in the measured spectroscopic redshift. We conclude that the detected line at 97.31 GHz is $^{12}$CO(3–2) from the lensed background source at $z = 2.553$. Full details of the LMT observations towards 9io9, together with a more complete analysis of the AzTEC and RSR data, are presented in Sanchez-Argüelles et al. (2015 in preparation) and Harrington et al. (2015 in preparation).

### 2.3 Near-infrared spectroscopy

We observed 9io9 with the Infrared Camera and Spectrograph (IRCS, Kobayashi et al. 2000) instrument on the Subaru telescope on the night of 15 January 2014 UT as a back-up target for the scheduled observing program[3]. The instrument was setup with the HK low-resolution grism covering the wavelength range 1.4–2.45$\mu$m. We used the 0.9″ wide slit of the 'Reflective 3W' slit-mask, resulting in a spectral resolving power of $R \approx 95$ at the central wavelength. The entrance slit was set at a position angle of 155 degrees east of north to align it to the bright, western half of the red ring. We observed 9io9 using the nod-on-slit technique with slit positions separated by 7″. We obtained 12 exposures for a total of 1 hour on source. Calibration data comprised of Argon lamp exposures for wavelength calibration, and Halogen lamp exposures for flatfield calibration.

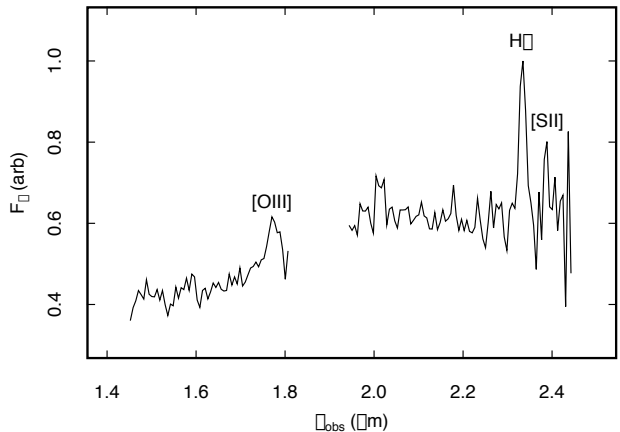

**Figure 4.** Observed near-infrared HK spectrum of the primary arc of 9io9 taken with IRCS on Subaru. The low resolution spectrum shows the clear rest-frame optical emission lines of the unresolved [O III]$\lambda$4959,$\lambda$5007 doublet, H$\alpha$+[N II], and the [S II]$\lambda$6717,$\lambda$6731 doublet, confirming the redshift of the lensed galaxy at $z = 2.553$ (see §2.3).

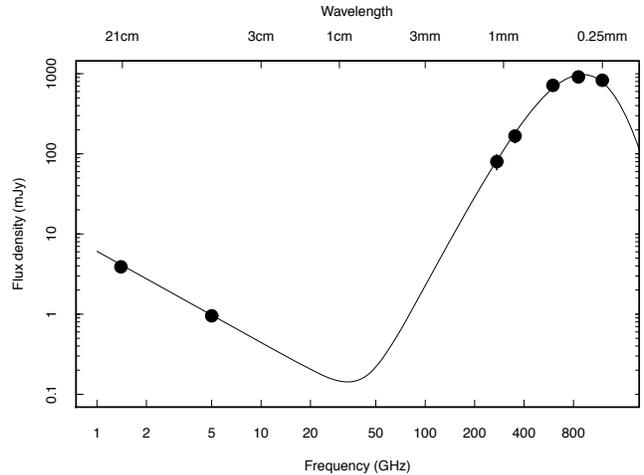

**Figure 5.** Observed radio-to-sub-millimetre spectral energy distribution (SED) of 9io9, showing the integrated flux density of the lensed galaxy (we assume no contribution from the lensing galaxy at these frequencies). The SED is well-modelled by a combination of modified blackbody with $T \approx 40$ K, describing the cold dust thermal emission, and a simple power law $\nu^{-\alpha}$ at $\nu < 50$ GHz, with spectral index $\alpha \approx 1.1$, describing radio synchrotron emission.

The data were reduced with the IRAF *twodspec* package, and standard procedures for flat-fielding and sky-subtraction for the nod-on-slit technique were used. The final data product is a two-dimensional spectrum of 9io9 with a regular sampling along the spatial and spectral directions. To increase the signal-to-noise ratio we binned the data in spectral direction by a factor of 6× and in the spatial direction by a factor of 4×. The total integrated spectrum of the western arc of the red ring of 9io9 is shown in Figure 4. The redshifted nebular emission lines of H$\alpha$+[N II] and (unresolved) [O III]$\lambda$4959,$\lambda$5007 and [S II]$\lambda$6717,$\lambda$6731 doublets are detected confirming the CO redshift of 2.553.

## 3 ANALYSIS

### 3.1 Spectral energy distribution

In Figure 5 we plot the (integrated) observed radio–sub-millimetre spectral energy distribution (SED) of 9io9. We model the sub-millimetre-millimetre photometry with a single temperature ($T$) modified blackbody

$$S_\nu \propto \frac{(1 - e^{-\tau(\nu)})\nu^3}{e^{h\nu/kT} - 1} \qquad (1)$$

where the optical depth $\tau(\nu) = (\nu/\nu_0)^\beta$ is normalised to 100$\mu$m (rest) and $\beta$ is the emissivity parameter. Since the redshift is accurately known, we allow $\beta$ and $T$ to be free parameters. The best fitting parameters are $\beta = 2.2 \pm 0.4$, $T = (38 \pm 4)$ K which provides a good model of the observed emission over 250–1100$\mu$m. The integrated (rest-frame) 8–1000$\mu$m luminosity is $\mu L_{IR} = (1.3 \pm 0.1) \times 10^{14} L_\odot$, where $\mu$ is the lensing magnification. Note that this might underestimate the total infrared luminosity since we have not modelled mid-infrared emission from warmer dust components. We fit the radio portion of the SED with a power law $S_\nu \propto \nu^{-\alpha}$, with $\alpha = 1.11 \pm 0.05$. The combined model SED is shown in Figure 5.

### 3.2 Lens modelling

The system shows a primary lensing galaxy G1 and a secondary satellite galaxy G2, probably at the same redshift ($z = 0.206$, see Fig. 1, 2). We choose an isothermal ellipsoid for the density profile of G1. Since G2 is located close to, and likely part of, the lens system, we model its density profile with an additional isothermal sphere. The parameters of the model are: positions and critical radius ($b$) for G1 and G2, ellipticity ($\epsilon$) and position angle (PA) for G1 ($\theta_\epsilon$), together with an external shear magnitude ($\gamma$) and PA ($\theta_\gamma$). We use a Gaussian prior ($\epsilon = 0.1 \pm 0.1$) on the ellipticity of G1 and a uniform prior on the position of G2 obtained from the optical light profile fitting of the galaxies.

The lens mass model is optimized to simultaneously fit the extended images from the $K_s$-band image (seeing $\theta = (0.8 \pm 0.1)''$) and the VLA 5 GHz data where the lensed arc and the counter image are detected with sufficient signal-to-noise. **The absolute astrometric calibration of the radio images is better than 100 mas, and we cannot improve the relative alignment of the near-infrared and VLA images. Furthermore, we acknowledge that** setting the source model to be coincident in the near-infrared and radio is an assumption that does not take into account the possibility that these have different source plane morphologies. However, it is necessary to make this assumption with the current data in order to reduce free parameters and degeneracies in the modelling. An annulus centered on the lens enclosing the arc and the counter image is used as a constraint when optimizing the lens model. We model the background source in the near-infrared with a Sérsic profile (flux, position, ellipticity and position angle, effective radius and Sérsic index) and in the radio with a Gaussian profile. We set the sources at both wavelengths to be coincident as we expect them to be physically associated.

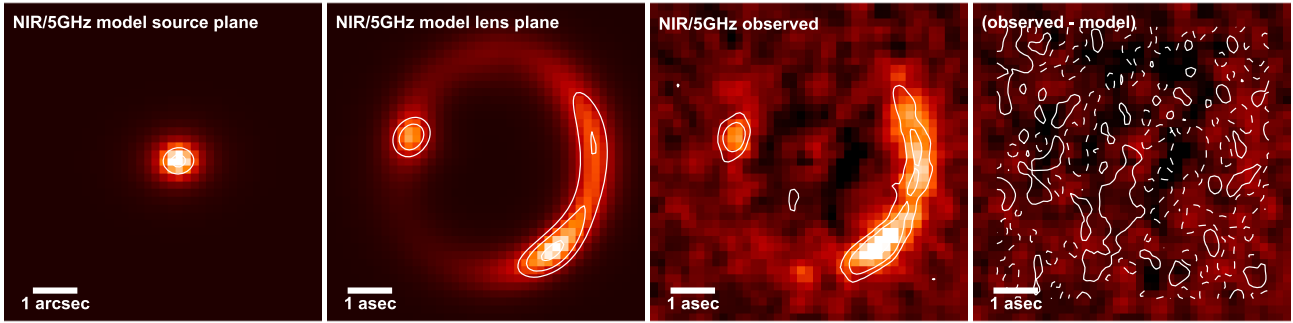

**Figure 6.** Mass modelling results based on joint fitting of the near-infrared imaging (colour map) and 5 GHz imaging (shown as contours at levels of $0.0033 \times (3,9,27)$ mJy beam$^{-1}$). (left) source plane model (centre left) the reconstructed image in the lens plane with the lensing galaxies G1 and G2 subtracted, (centre right) observed data linearly scaled between $-3$–$10\sigma$, (right) residual (observed$-$model). The residual colour scale (near-infrared) is the same as the previous panel and the contours are at levels of $0.0033 \times (-1,1)$ mJy beam$^{-1}$. Images are orientated north up, east left.

To construct the mass model, we use *gravlens* (Keeton 2011) and a Markov Chain Monte Carlo (MCMC) simulation for the fitting procedure (using the Python module *emcee*, Foreman-Mackey et al. 2013) to optimize the parameters and explore degeneracies in the model. Initially, the position of G1 is held fixed during optimization and after a reasonable model is found the position is allowed to be free. The source plane model, model lensed images and residuals (for both wavelengths) are shown in Figure 6. The magnification in the near-infrared is $\mu_{\rm NIR} \approx 11$ and in the radio is $\mu_{\rm 5\,GHz} \approx 13$, the unlensed magnitude of the source is $K_{\rm s} = 20.75$ mag and $S_{\rm 5\,GHz} = 1.2$ $\mu$Jy beam$^{-1}$. The parameters of the best fit lens model are summarized in Table 1. The critical radii from Table 1 can be taken with the lens source redshifts to calculate the lens velocity dispersion for G1 and G2: $\sigma_{\rm G1} = 316 \pm 1$ km s$^{-1}$ and $\sigma_{\rm G2} = 80 \pm 4$ km s$^{-1}$.

Although the eMERLIN data are of somewhat lower signal-to-noise, their higher resolution allows at least preliminary investigation of the radio source structure. Using the galaxy mass model established above for the $K_{\rm s}$-band and VLA imaging, we can attempt to reconstruct the eMERLIN data using a parametrised Gaussian fit to represent the source. The results are shown in Figure 7. In general, the source reconstruction is subject to considerable degeneracy because the main constraint provided by the lensing concerns the source distribution around the high-magnification region corresponding to the caustic running across it. Nevertheless, the data indicate a double rather than a single Gaussian fit, if we believe the structure at the southern end of the arc which is difficult to reproduce with a single component. The best-fitting model is subject to some degeneracies with the existing data, which can be investigated using a standard MCMC procedure. Specifically, the relative position of the centres of the components has an error approximately equal to their separation in the best-fitting model, with the more compact component contributing most of the positional error. The parameters describing the compact component, including its precise position, are uncertain apart from its size, which needs to be below approximately 30 mas, or 250 pc in the source plane. The more extended component is preferred to be elliptical ($b/a = 0.27^{+0.15}_{-0.10}$ with PA $15 \pm 24$ degrees), and together the radio structure is reasonably consistent with that of a small radio core and jet, although higher signal-to-noise data would be needed to make a more definitive statement. We defer this, together with a more systematic source reconstruction using other methods to a future work.

**Table 1.** Best-fitting parameters describing the lens system comprised of components G1 and G2. Positions are given relative to the coordinates ($02^{\rm h}09^{\rm m}41.268^{\rm s}$, $+00^\circ 15' 58.59''$). Note that there is a degeneracy between the ellipticity of G1 and the shear magnitude and similarly, between the mass (or the critical radius) of G1 and G2. Uncertainties reflect the 16th and 84th percentile range derived from the MCMC fitting of the mass model.

| Parameter | G1 | G2 |
|---|---|---|
| x (mas) | $-94^{+9}_{-13}$ | $377^{+17}_{-32}$ |
| y (mas) | $126^{+4}_{-10}$ | $2332^{+255}_{-2}$ |
| | | |
| Critical radius ($b$, arcsec) | $2.48^{+0.02}_{-0.01}$ | $0.16^{+0.02}_{-0.01}$ |
| Ellipticity ($\epsilon$) | $0.12 \pm 0.02$ | — |
| PA ($\theta_\epsilon$, deg) | $84^{+3}_{-1}$ | — |
| | | |
| Shear magnitude ($\gamma$) | $0.064^{+0.005}_{-0.004}$ | — |
| Shear PA ($\theta_\gamma$, deg) | $83^{+1}_{-2}$ | — |

## 4 DISCUSSION

By simultaneously fitting the $K_{\rm s}$-band and 5 GHz imaging we have derived a lens model that allows us to determine a magnification factor of approximately $10\times$. The magnification-corrected infrared luminosity of 9io9 is $L_{\rm IR} \approx 10^{13} L_\odot$, ranking it amongst the most luminous galaxies in the Universe, in the 'hyperluminous' class. If the entire infrared luminosity budget is dominated by star formation (i.e. re-processed ultraviolet emission from massive stars), then the implied star formation rate of 9io9 is $\dot{M}_\star \approx 2500\,M_\odot\,\rm yr^{-1}$ (Kennicutt & Evans 2012, assuming a Chabrier initial mass function), a thousand times the star formation rate of the Milky Way.

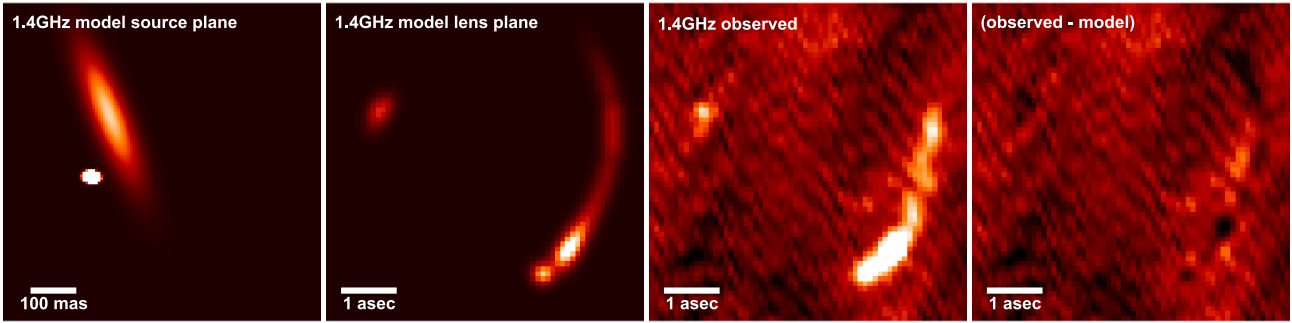

**Figure 7.** (left) the reconstructed 1.4 GHz source model, using a two-Gaussian fit and the lens model described in §3.2. Note that the relative positions of the compact and extended components are not well constrained in the current model. (centre left) the reconstructed image in the lens plane, (centre right) eMERLIN observed data linearly scaled between -5–10$\sigma$, (right) residual (observed−model). The residual colour scale is the same as the previous panel. Note the scale bar for the source plane is 100 mas, corresponding to 820 pc in physical projection. **With the exception of the first panel, images are centred on the position of the lensing galaxy G1 (Figure 1). The source plane image is slightly offset from centre for illustration purposes.** Images are orientated north up, east left.

Given the radio flux density and reconstructed radio morphology described above, it is likely that 9io9 contains an AGN that contributes non-negligibly to the energy output, and therefore $\dot{M}_\star$ should be taken as an upper limit, with the actual star formation rate scaling like $\gamma L_{\mathrm{IR}}$, where $\gamma \approx 0.6$ for starburst/AGN composite systems at similar epoch with comparable $L_{\mathrm{IR}}$, determined through mid-infrared diagnostics (Coppin et al. 2010). Even with a 40% correction for AGN contribution to $L_{\mathrm{IR}}$, the implied star formation rate is still remarkably high, at over a thousand solar masses per year.

In the absence of other diagnostics (such as the observation of broad lines, or emission line ratios such as [OIII]/H$\beta$), the infrared-to-radio luminosity ratio can be used as a basic diagnostic for the presence of an active galactic nucleus, described by the parameter

$$q = \log_{10}\left(\frac{L_{\mathrm{IR}}/3.75 \times 10^{12}\,\mathrm{W\,m^{-2}}}{L_{1.4\,\mathrm{GHz}}/\mathrm{W\,m^{-2}\,Hz^{-1}}}\right) \qquad (2)$$

with the monochromatic radio luminosity $K$-corrected to the rest-frame 1.4 GHz assuming the measured $\nu^{-1.1}$ spectral slope. Star-forming galaxies have a rather tight distribution of $q$, with $\langle q \rangle \approx$ 2.5 and scatter of approximately $\sigma_q \approx 0.25$ (e.g. Ivison et al. 2010). The radio–far-infrared correlation is thought to arise from two linked processes associated with massive star formation: (a) cosmic-ray electrons from supernovae that emit synchrotron radiation as they are accelerated in magnetic fields, leaking out of dust-enshrouded HII regions, and (b) infrared emission originating from dust-reprocessed ultraviolet photons emitted by young massive stars. In galaxies hosting AGN, synchrotron emission associated with black hole accretion can contribute to the radio flux, resulting in lower $q$. In this case, the magnification-corrected radio luminosity is $L_{1.4\,\mathrm{GHz}} = (1.7 \pm 0.1) \times 10^{25}\,\mathrm{W\,Hz^{-1}}$, implying $q = 1.8 \pm 0.1$ (here we have assumed the same magnification correction for the radio and infrared emission, $\mu = 13$). Thus, 9io9 is not consistent with a pure starburst, with significant nuclear activity associated with the growth of the central black hole contributing to the radio emission.

The current data allow us to postulate, with a reasonable degree of confidence, the basic nature of the radio source. There are some hints that the radio spectral index is flatter in the core component, $\alpha \approx 0.2$ steepening to $\alpha \approx 1$ along the ring. This would be consistent with the classic picture of optically thick emission due to synchrotron self-absorption around a compact, dense nuclear source. The steepening spectrum could indicate aging of the rela-tivistic electron population (which lose their energy as $E^2$) along a possible jet, but a more detailed analysis that takes into account the different visibilities sampled by eMERLIN and VLA, improved sampling of the radio spectrum and a more accurate lens model will be needed to investigate this. Considering the reconstructed source plane morphology, with compact core and potential jet/lobe structure, the radio power of 9io9 places it in the Fanaroff-Riley II class of radio galaxies (FRII, $L_{1.4\,\mathrm{GHz}} > 10^{25}\,\mathrm{W\,Hz^{-1}}$).

# 5 SUMMARY

Through a citizen science project – SPACE WARPS – we have discovered an extremely luminous galaxy at $z = 2.5$, close to the peak in stellar mass and black hole growth in the Universe, lensed by a foreground **Luminous Red Galaxy, possibly part of a small group (given the large Einstein radius compared to typical single-galaxy lenses)** at $z = 0.2$. The lensed source, dubbed 9io9, appears as a red Einstein ring, prominent in the near-infrared, but very faint in the optical (and thus would have been overlooked in an optical survey), characteristic of either a heavily dust-obscured or predominantly old stellar population. This lensed system would have been overlooked in an optical survey. Archival and subsequent follow-up observations revealed that 9io9 is extraordinarily bright in the sub-millimetre, with a flux density approaching 1 Jy at the peak of the thermal dust emission. We have resolved the lens at two radio frequencies, with sub-arcsecond eMERLIN and VLA mapping at 1.4 GHz and 5 GHz respectively, revealing radio emission from the source tracing the near-infrared light. The strong radio and infrared emission indicates a galaxy undergoing an episode of rapid growth, either in an intense starburst episode, through significant matter accretion onto a central supermassive black hole, or both.

A plausible evolutionary scenario for 9io9 is that the galaxy is destined to evolve into a massive ($M_\star \approx 10^{11-12}\,M_\odot$) elliptical galaxy by $z = 0$, perhaps located at the centre of a large group or cluster of galaxies. Seen when the Universe was less than 3 Gyr old, we are gaining an insight into the major growth phase of this massive galaxy, with gravitational lensing providing a unique window onto spatial scales not possible for an equivalent non-lensed field source. The fact that high luminosity sources such as this are intrinsically rare even in the absence of lensing highlights the benefits that citizen science can offer in the mining of (ever growing) astronomical imaging archives, albeit at the cost of rather complex selection functions for the objects found. We plan further follow-up of 9io9 to improve our understanding of the source. Particularly exciting is the possibility that we could detect and monitor variability in the source, which – if associated with the compact nuclear core – could be measured with cadences of days, weeks or months.

## ACKNOWLEDGEMENTS

We thank the anonymous referee, whose comments improved the clarity of this work. This result would not have been possible without the help of thousands of volunteers participating in the project through BBC *Stargazing Live!* and the *Zooniverse*. We thank all of the Citizen Scientists who participated in SPACE WARPS. The full list of their *Zooniverse* usernames can be found at the following URL: http://spacewarps.org/#/projects/VICS82/summary. We thank Zbigniew 'Zbish' Chetnik, one of the volunteers who classified 9io9, who appeared on the *Stargazing Live!* programme to discuss the object. It is our pleasure to thank the BBC Stargazing Live production team for their efforts in making SpaceWarps a success: Liz Bonnin, Mark Thompson (presenters), Ross Fiddes, Gavin Hesketh, Amy Hewer, Ben Lucas, Abi Rose, Adam Toth (production team), Holly Salter, Lizzie Search, Lauren Webber (production co-ordinators), Clare Huggins (production manager), Laura Davey (production executive), Alex Dackevych, Simon Mackie (web team), Ayo Ajibewa, Faran Ismailpour, Ivan Lazic, Kiki Lawrence, Keaton Stone (researchers), Alastair Duncan (assistant producers), Tom Hewitson, Alice Jones, Penny Palmer, Sarah Sarkhel (VT directors), Chris Parkin, Paul Vanezis (producers), Ian Russell (director), Helen Thomas (executive producer), Paul King (series producer).

J.E.G. is supported by a Royal Society University Research Fellowship. A.M. is supported by World Premier International Research Center Initiative (WPIInitiative), MEXT, Japan. A.M. also acknowledges the support of the Japan Society for Promotion of Science (JSPS) fellowship. e-MERLIN is a National Facility operated by the Univer- sity of Manchester at Jodrell Bank Observatory on behalf of the UK Science and Technology Facilities Council (STFC). A.V. acknowledges support from a Leverhulme Trust Research Fellowship. R.J.I. acknowledges support from the European Research Council in the form of Advanced Grant, 321302, COSMICISM. J-P.K. acknowledges support from the ERC advanced grant LIDA and from CNRS. M.M. is partially supported by CNPq and FAPERJ. e-MERLIN is a National Facility operated by the University of Manchester at Jodrell Bank Observatory on behalf of the STFC. Based in part on data collected at Subaru Telescope, which is operated by the National Astronomical Observatory of Japan. This work is based on observations carried out with the VLA. The NRAO is a facility of the NSF operated under cooperative agreement by Associated Universities, Inc. The James Clerk Maxwell Telescope has historically been operated by the Joint Astronomy Centre on behalf of the STFC, the National Research Council of Canada and the Netherlands Organisation for Scientific Research. Additional funds for the construction of SCUBA-2 were provided by the Canada Foundation for Innovation. This research has made use of the NASA/IPAC Extragalactic Database (NED) which is operated by the Jet Propulsion Laboratory, California Institute of Technology, under contract with the National Aeronautics and Space Administration. Based on observations obtained with WIRCam — a joint project of Taiwan, Korea, Canada, France, and the Canada-France-Hawaii Telescope (CFHT) — and MegaPrime/MegaCam — a joint project of CFHT and CEA/DAPNIA — at CFHT and VIRCAM at the VISTA/ESO Telescope at the Paranal Observatory. CFHT is operated by the National Research Council (NRC) of Canada, the Institut National des Sciences de l'Univers of the Centre National de la Recherche Scientifique (CNRS) of France, and the University of Hawaii. The Brazilian partnership on CFHT is managed by the Laboratório Nacional de Astrofísica (LNA). We thank Terapix for the VISTA data reduction. We thank the CFHTLens team for their pipeline development on which the CS82 reductions were based. We thank the Mexican Science and Technology Funding Agency, CONACYT (Consejo Nacional de Ciencia y TecnologÃ­a) during the construction and early operational phase of the Large Millimeter Telescope Alfonso Serrano, as well as support from the the US National Science Foundation via the University Radio Observatory program, the Instituto Nacional de Astrofísica, Optica y Electronica (INAOE) and the University of Massachusetts (UMASS). The UMASS LMT group acknowledges support from NSF URO and ATI grants (AST-0096854 and AST-0704966) for the LMT project and the construction of the RSR.